\documentclass[twocolumn,showpacs,preprintnumbers,amsmath,amssymb,superscriptaddress, natbib]{revtex4-1}
\usepackage{amssymb}
\usepackage{graphicx}
\usepackage{dcolumn}
\usepackage{bm}
\usepackage{times}
\usepackage{hyperref}
\usepackage{color}
\usepackage{textcomp}
\newcommand{\SLFCO}{SrLaFeCoO$_6$}
\newcommand{\SFCO}{Sr$_{2}$FeCoO$_{6}$}
\newcommand{\LFCO}{La$_{2}$FeCoO$_{6}$}
\newcommand{\SLFCOx}{La$_{2-x}$Sr$_x$FeCoO$_6$}

\begin{document}
%

\title{Contrasting the magnetism in La$_{2-x}$Sr$_x$FeCoO$_6$ ($x$ = 0, 1, 2) double
	perovskites: the role of electronic and cationic disorder}
\author{G. R. Haripriya}
\affiliation{Low Temperature Physics Laboratory, Department of Physics, Indian Institute of Technology Madras, Chennai, India}
\author{C. M. N. Kumar}
\email{naveen.chogondahalli@ifp.tuwien.ac.at }
\affiliation{Institut f\"{u}r Festk\"{o}rperphysik, TU Wien, Wiedner Hauptstr. 8-10/138, 1040 Wien, Austria}
\author{R. Pradheesh}
\affiliation{Low Temperature Physics Laboratory, Department of Physics, Indian Institute of Technology Madras, Chennai, India}
\author{L. M. Martinez}
\affiliation{Department of Physics, 500 W. University Ave, University of Texas at El Paso, El Paso, TX 79968, USA}
\author{C. L. Saiz}
\affiliation{Department of Physics, 500 W. University Ave, University of Texas at El Paso, El Paso, TX 79968, USA}
\author{S. R. Singamaneni}
\affiliation{Department of Physics, 500 W. University Ave, University of Texas at El Paso, El Paso, TX 79968, USA}
\author{T. Chatterji}
\affiliation{Institut Laue-Langevin, BP 156, 38042 Grenoble Cedex 9, France}
\author{V. Sankaranarayanan}
\affiliation{Low Temperature Physics Laboratory, Department of Physics, Indian Institute of Technology Madras, Chennai, India}
\author{K. Sethupathi}
\affiliation{Low Temperature Physics Laboratory, Department of Physics, Indian Institute of Technology Madras, Chennai, India}
\author{B. Kiefer}
\affiliation{Department of Physics, New Mexico State University, Las Cruces, New Mexico 88011,USA}
\author{H. S. Nair}
\email{hnair@utep.edu}
\affiliation{Department of Physics, 500 W. University Ave, University of Texas at El Paso, El Paso, TX 79968, USA}
\date{\today} 

\begin{abstract}
The magnetism of the double perovskite
compounds \SLFCOx\ ($x$ = 0, 1, 2) are contrasted using magnetization, 
neutron diffraction and electron paramagnetic resonance with the 
support from density functional theory calculations. 
\LFCO\ is identified as a long-range ordered antiferromagnet displaying
a near-room temperature transition at $T_N$ = 270~K, accompanied by
a low temperature structural phase transition at $T_S$ = 200~K.
The structural phase transformation at $T_S$ occurs from 
$R\overline{3}c$ at 300~K to $Pnma$ at 200~K. 
The density functional theory calculations support an
insulating non-compensated AFM structure.
The long-range ordered magnetism of \LFCO\ transforms to
short-range glassy magnetism as La is replaced with Sr in the other two compounds.
The magnetism of \LFCO\ is differentiated from the 
non-equilibrium glassy features of \SFCO\ and \SLFCO\
using the {\em cooling-and-heating-in-unequal-fields} (CHUF) magnetization protocols.
This contransting magnetism in the \SLFCOx\ series is evidenced in electron paramegnetic
resonance studies.
The electronic density-of-states estimated using the density functional
theory calculations contrast the insulating feature of \LFCO\ from the
metallic nature of \SFCO.
From the present suite of experimental and computational results on \SLFCOx, 
it emerges that the electronic degrees of freedom, along with antisite disorder, 
play an important role in controlling the magnetism observed in double perovskites.
\end{abstract}
\maketitle

\section{Introduction}
\label{section_intro}
\indent 
Double perovskites Sr$_2BB'$O$_6$, where $B/B'$ are transition metal elements, attracted 
attention due to the observation of large magnetoresistance in the case of
$B/B'$ = Fe/Mo \cite{kobayashi1998room,sarma2000magnetoresistance,sarma2001new}. 
The cation-ordered Sr$_2$FeMoO$_6$ was reported to show room temperature, low-field 
magnetoresistance (MR) with the striking feature of scaling of MR with the square 
of spin-polarization of carriers, $(M/M_s)^2$, where $M_s$ is the saturation 
magnetization \cite{kobayashi1998room}. 
This suggested the potential for spintronics and giant magnetoresistance applications in an ideally 
ferromagnetic double perovskite lattice, which motivated experimental 
studies connected to the low-field MR in Sr$_2$FeMoO$_6$ \cite{sarma2000magnetoresistance,navarro2001antisite}. 
However, double perovskites prepared at high temperatures in laboratories
suffer from antisite disorder on the $B/B'$ site. 
This leads to the disruption of the $B^{2+} \textendash$ O $\textendash B'^{4+}$ 
magnetic exchange paths and consequent weakening of ferromagnetism 
predicted by the Goodenough-Kanamori rules for an ordered cation arrangement of cations 
\cite{kanamori_10_87_1959,goodenough_100_564_1955}. 
Antisite disorder has a strong bearing on the magnetic and the transport 
behaviour of the double perovskites. 
Significant differences in MR at low temperature (4.2~K) was reported in the case of 
ordered (degree of Fe/Mo ordering = 91$\%$), versus the disordered 
(degree of Fe/Mo ordering = 31$\%$) Sr$_2$FeMoO$_6$. 
Though the Fe/Mo-based Sr$_2BB'$O$_6$ was studied in detail for its
magnetoresistive properties, less attention was paid to the Fe/Co based Sr$_2BB'$O$_6$ 
compounds which offer the possibility of tuning the structural, valence and spin-state 
parameters connected to the magnetic behaviour. 
This is particularly possible due to the presence of Co, which can adopt 
low-spin (LS), high-spin (HS) or intermediate-spin (IS) states depending 
on the valence that is stabilized in a particular double perovskite structure. 
The orbital degrees of freedom and consequently the spin-orbit 
coupling effects attain importance in this case. \\
\indent 
It is, hence, understood that the crystallographic antisite disorder has a significant
impact on the magnetism of double perovskites.
Another important structural detail that has a significant bearing is the
distortions of the metal-oxygen octahedra that constitute the perovskite.
The ideal perovskite $AMX_3$ structure adopts highly symmetric cubic space
group $Pm\overline{3}m$ where the $A$ cation is surrounded by 12 $X$ anions and the $M$
cation by 6 $X$ anions.
The three-dimensional view of the perovskite structure is that of a corner-sharing 
$MX_6$ octahedra. 
Distortions, tilting or cation displacements in the octahedra lead to a deviation from the ideal 
cubic structure and can lead to low symmetry space groups like $P2_1/n$ or $I4/m$.
A convenient classification of how the tilts in the perovskite structure leads to different
space groups symmetries is provided by Glazer \cite{glazer1972classification}.
Using this system, a tilt in the octahedra is described by specifying the rotations of the 
octahedra about each of the three cartesian axes.
The $Pm\overline{3}m$ space group belongs to the tilt system $a^0a^0a^0$ 
and the rhomboehedral space group $R\overline{3}c$, $a^-a^-a^-$.
The rotation pattern of the orthorhombic $Pnma$ space group is determined by
two tilts which are $a^0a^0a^+$ and $a^-a^-a^0$.
The octahedral tilts and rotations are extremely important to single and layered perovskite 
compounds in bringing about novel type of ferroelectricity \cite{benedek2012polar}.
The double perovskite compounds are generally found to the adopt 
random, rock salt or layered structure types depending on the degree of $B$ cation
arrangement \cite{anderson1993b}.
Space groups $Pm\overline{3}m$ and $Pnma$ (random), $Fm\overline{3}m$, $P2_1/n$
(rock salt) and $P2_1/m$ (layered) were predicted based on this \cite{anderson1993b}. \\
\indent 
The role of cation disorder in the crystal structure and magnetism of the 
Sr$_2BB'$O$_6$ compound \SFCO\ (SFCO) was investigated by some of us \cite{pradheesh}. 
SFCO was seen to adopt the tetragonal space group $I4/m$ with the lattice parameters, $a$ = 5.4609(2) {\AA} and $c$ = 7.7113(7) {\AA}; which is about 2$\%$ reduced in the $a$ and $b$ compared to those of Sr$_2$FeMoO$_6$. 
The magnetic ground state is identified as a canonical spin glass with a spin freezing temperature, 
$T_g \approx$ 75~K, \cite{pradheesh} which is quite different from the ferrimagnetic ground state of 
Sr$_2$FeMoO$_6$ with a $T_c$ in the range 410 $\textendash$ 450~K \cite{patterson1963magnetic}.
Albeit the differences in the magnetic ground state and the lattice parameters, SFCO displays 
large magnetoresistance of 63$\%$ at 14~K in 12~T \cite{pradheesh2012large}. 
Strong antisite disorder was observed in SFCO along with the presence of mixed valence 
states for Co.
The disorder effect and mixed valence in SFCO gave rise to not only the spin glass 
magnetism, but also to large magnetoresistance derived from the spin scattering 
of the carriers localized by the magnetic moments in the spin glass state. 
Additionally, it also lead to the development of exchange bias \cite{pradheesh_exchangebias}. 
Upon replacement of Sr with La in the case of \SLFCO (SLFCO), features of a magnetic glass 
were observed \cite{pradheesh2017magneticglass}. 
The magnetization of SLFCO showed an anomaly at $T_\mathrm{a1} \approx$ 75~K.
Despite the non-equilibrium metastable magnetic state, significant magnetoresistance of 
about 47$\%$ was observed in SLFCO at 5~K in 8~T \cite{pradheesh2012large}. 
With the addition of La, a significant change in the crystal structure was the stabilization of monoclinic 
space group $P2_1/n$.
Although the monoclinic structure is amenable to perfect ordering of Fe and Co in two different Wyckoff 
positions $2c$ and $2d$, a high degree of disorder ($\approx$ 90$\%$) was observed in SLFCO.  \\
\indent 
The present paper extends the work on SFCO and SLFCO to the crystal structure and 
magnetism of \LFCO\ (LFCO).
Using the experimental tools of magnetization, neutron diffraction and electron 
paramagnetic resonance (EPR) we study the structure and magnetism in LFCO
and compare it with that of SFCO and SLFCO.
Density functional theory (DFT) calculations on all the three compounds support our
experimental findings.
It is seen that LFCO develops magnetic long-range order at significantly high temperatures ($\approx$ 270~K) and subsequently undergoes a structural phase transition at 200~K.
The magnetism in LFCO is opposed to that of SFCO and SLFCO, which are seen to be 
magnetically disordered below $\approx$ 75~K.

\section{Methods}
\subsection{Experimental techniques}
Polycrystalline samples of \LFCO\ were prepared following sol-gel 
method as described in Reference [\onlinecite{pradheesh}], which explains the preparation of SFCO.
For the present work, LFCO and SLFCO were prepared using a similar synthesis method. 
The synthesized compounds were first analyzed using powder X-ray 
diffraction to check phase purity and crystal structure. 
Magnetic measurements were carried out on pressed pellets in a 
Magnetic Property Measurement system SQUID Vibrating Sample 
Magnetometer (MPMS-SVSM) in the temperature range, 5 $\textendash$ 350~K 
and magnetic field $\pm$7~T. 
Zero field cooling (ZFC), field-cooled warming (FCW) and field cooled cooling (FCC) protocols
were used to measure dc magnetization.
Neutron powder diffraction experiments on LFCO and SLFCO were performed at 
WISH (Rutherford Appleton Laboratory, UK) \cite{wish}. 
Roughly 8~g of well-characterized powder sample was used for each neutron experiment. 
The diffraction data was analyzed using Fullprof Suite of programs 
\cite{fullprof} for Rietveld refinements and the software SARA$h$ \cite{sarah_wills} 
was used for the analysis of magnetic structure using representation analysis. 
EPR data were recorded on a Bruker EMX Plus X-band ($\approx$ 9.43~GHz) 
spectrometer, equipped with a high sensitivity probe head. 
A ColdEdge$\textsuperscript{TM}$ ER 4112HV In-Cavity Cryo-Free VT 
system connected with an Oxford temperature controller was used for low 
temperature measurements.

\subsection{Computational methods}
\indent
We first present the results of the parameter-free first-principles density 
functional theory computations \cite{hohenberg1964inhomogeneous, kohn1965self} 
to elucidate the structure and magnetism in the series of three compounds \SFCO, \SLFCO\ and \LFCO.
Our calculations take into account the experimental low temperature crystal structure details and hence are more reliable than the previous reports. All computations were performed for the relevant low temperature structures, obtained from our neutron experiments. All computations were performed for fixed lattice and positions.
We determined electronic and magnetic properties for SFCO, SLFCO and LFCO, 
for fixed structure, neglecting relaxation effects of any crystallographically allowed degrees of freedom. 
All computations were performed with the 3D planewave software package VASP \cite{kresse1993ab, kresse1996g, kresse1996prb} with the projector-augmented wave method \cite{blochl1994projector, kresse1999ultrasoft} with the PBE-GGA exchange correlation functional \cite{perdew1996generalized} 
and included on-site Coulomb interactions (DFT + U) \cite{dudarev1998electron}. 
U = 5.0~eV for Fe and Co, $E_\mathrm{cut}$ = 400~eV and a $k$-point spacing of 0.25 were used 
for all computations, similar to the previous work on LFCO \cite{fuh2015electronic} 
and related compounds \cite{wu2011theoretical,pruneda2008structural, yang1999influence}.
The $k$-point density ensured that it was sampled homogeneously across different crystal structures, 
facilitating a comparison of computed properties. 
For Sr, La, Fe, Co, and O, we considered explicitly 5$s^2$4$p^6$4$s^2$, 
5$d^1$6$s^2$5$p^6$5$s^2$, 3$d^6$4$s^1$, 3$d^7$4$s^2$, and 2$s^2$2$p^4$ as shells 
in our computations, respectively. 
In order to explore the effect of spin-lattice coupling and the robustness of the electronic structure in the 
vicinity of the Fermi energy, we included spin-orbit coupling in the most stable configurations and in order to resolve better the small effects of spin-orbit coupling we used a denser $\Gamma\textendash$centered $k\textendash$point grid with a spacing of 0.15.

\section{Results}
\subsection{Density functional theory of \SLFCOx}
\subsubsection{\LFCO}
\begin{figure}[!t]
	\centering
	\includegraphics[scale=1.05]{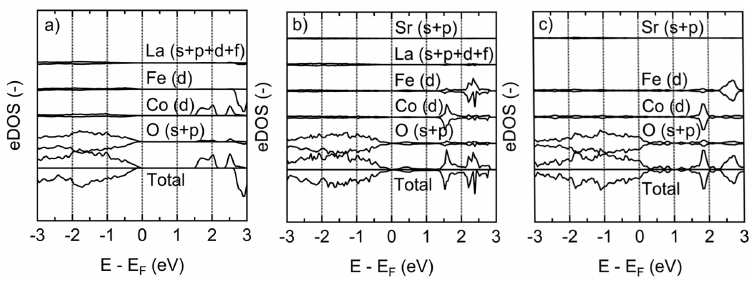}
	\caption{\label{fig_dft} 
		The total electronic density of states (eDOS) of the ground state for
		(a) LFCO (G-structure), (b) SLFCO (C-structure) and (c) SFCO (G-structure) obtained through density functional theory calculations.}
\end{figure}

he electronic density of states (eDOS) for all the three compositions of \SLFCOx are presented in Fig~\ref{fig_dft}. 
We tested the oxidation states of all three compounds with different cation distributions denoted as A, C, and G, adopting the labeling of magnetic structures in perovskites \cite{wollan1955neutron}.
If the two Fe ions are in the same plane perpendicular to the long axis, we refer to an A-structured arrangement (4 Fe and 2 Co nearest neighbors); if the two Fe ions are on a line parallel to the long axis, they form a C-structured arrangement (2 Fe and 4 Co nearest neighbors); if all nearest-neighbors are of opposite type, they are labeled as G-structured arrangement (0 equal and 6 non-equal neighbors).

In the case of \LFCO, total energy differences between A, C, and G 
transition metal arrangements are less than 9.0~meV/5 atoms, 
consistent with strong (Fe,Co) antisite disorder. 
This is approximately an order of magnitude smaller in energy 
than computed for Ba$_x$Sr$_{1-x}$Co$_y$Fe$_{1-y}$O$_{3-\delta}$, 
where a single (Fe,Co) exchange requires 80~meV \cite{kuklja2012intrinsic}. 
The ground state is G-structure and insulating. 
The computed magnetic structure corresponds to non-compensated AFM with
site projected magnetic moments of
+4.2, +4.2, -3.1, -3.1~$\mu_\mathrm B$, 
for Fe and Co (Fig.~\ref{fig_dft} (a)). 
The magnetic moments are consistent with Fe$^{2+}$(HS) 
and Co$^{4+}$(IS) charge assignments, as reported before \cite{fuh2015electronic}, 
except that our magnetic ground state is not FM, 
but non-compensated AFM as described recently \cite{wang2018tunable}. 
All structure/oxidation states/magnetic moment arrangements 
converged to the same oxidation state, described above. 
Thus, no valence disorder is required to explain the 
magnetic ground state of LFCO. 
Similarly, A-, and C- structured transition metal arrays form excited states.
They are AFM with magnetic moments of +4.2, -4.2, -3.1, +3.1~$\mu_\mathrm B$, and 9~meV/5 atoms and 3~meV/5 atoms higher than the predicted to be non-compensated AFM.
This near-degeneracy of states may explain the small hysteresis in the 
magnetization curve of LFCO (Fig.~\ref{fig_mag_lfco} (b)). 
A-, C-, G-plaquettes are randomly generated at the 
high synthesis temperature ($T >$ 1000$^\circ$ C) and form the structural template 
for LFCO. 
If so, the magnetic state is likely a superposition of non-compensated AFM and AFM structure at low temperature.
Moreover, CHUF2 observations (presented in Section~\ref{section_chuf}) suggest 
unsaturated magnetic moments, in general agreement with our computed results. 
The small energy differences do suggest that spin-orbit coupling may influence 
the magnetic structure. 
Our results show that energy differences between different spin orientations are 1~meV/5 atoms. We find that spins sub-parallel to the 
[001] direction are energetically most favorable, followed by [010] and [111] 
magnetization directions. 
The orbital and spin moments for the most stable 
[001] magnetic structure are parallel for Fe, as expected from Hund’s rules 
for less than half-filled electron shells (3$d^4$). 
Interestingly, we find an orbital moment for Co, suggesting that Co$^{4+}$ is not in a high-spin state (3$d^5$). 
However, the orbital moments are at least two orders of magnitude 
smaller than the spin magnetic moments, but support a small canting confined to the 
$bc$-plane, consistent with our neutron scattering data (Section~\ref{section_neutron}). 
\subsubsection{\SLFCO}
\indent
In the case of SLFCO, the DFT ground state is AFM with a G-type transition 
and a C-type Sr, La arrangement and site-projected magnetic moments of +3.9, -3.9, -2.8, 
+2.8~$\mu_\mathrm B$, and semi-metallic electronic structure (Fig.~\ref{fig_dft} (b)). 
In an ionic picture $q$ (Fe) + $q$ (Co) = 7. 
The magnetic moments are consistent with Fe$^{2+}$ (HS) and Fe$^{4+}$ (HS), leaving Co in a +5 or +3 charge state, with even magnetic moments, in contrast to the computed moment. 
This discrepancy may be attributed to charge ordering, and the coexistence of Co$^{5+}$ (HS) and Co$^{3+}$ (IS) states suggesting the presence of mixed valence states \cite{pradheesh}. 
Moreover, we find that the second most stable phase is FM with the same cation arrangement as the ground state, but $\sim$1.5 meV/5 atoms less stable. 
The site projected magnetic moments are +4.0, +4.0, +2.0, +3.0~$\mu_\mathrm B$, for Fe, and Co, respectively. 
The magnetic moments of one of the Co atoms is predicted to decrease by $\sim$1~$\mu_\mathrm B$, as compared to the ground state, and consistent with a charge assignment of Co$^{3+}$(IS), and Co$^{5+}$(IS). 
The existence of a low lying FM state is in excellent agreement with the interpretation of our magnetometry (Section~\ref{section_chuf}) and our EPR results (Section~\ref{section_epr}).
\subsubsection{\SFCO} 
\indent
For SFCO, the DFT computations show that the ground state is an 
antiferromagnetic metal (Fig~\ref{fig_dft} (c)), with G-type transition 
metal ion and spin arrangement. 
The cell magnetic moment (20 atoms) is zero and the site projected magnetic moments are +3.7~$\mu_\mathrm B$, -3.7~$\mu_\mathrm B$, -2.9~$\mu_\mathrm B$, +2.9~$\mu_\mathrm B$ for Fe and Co respectively, in overall agreement with our experimental observations.
The second most stable state predicted is ferromagnetic with C-type arrangement, and ~4 meV/5 atoms less stable than the antiferromagnetic ground state, and a magnetic moment is ~10.6~$\mu_\mathrm B$/20 atoms. 
Given the small energy difference it is to be expected that G-type and C-type structural plaquettes can coexist at the high synthesis temperatures. 
More importantly, we note that the magnitude of the site-projected magnetic moments 
are 3.8~$\mu_\mathrm B$, 4.0~$\mu_\mathrm B$, 2.8~$\mu_\mathrm B$, and 2.9~$\mu_\mathrm B$, 
for the two Fe and Co atoms, respectively. 
Regardless of the initialized multiplet in the computations, the magnitude of the final spins was always within ~0.3~$\mu_\mathrm B$ of the ground state, supporting a common oxidation state. 
A consistent set of oxidation states is Fe$^{4+}$ (HS) and Co$^{4+}$ (IS), 
and similar to LFCO, the DFT results do not require valence state mixing for SFCO. 
Therefore, the DFT computations suggest that magnetic multiplets are energetically close and can coexist at low temperatures, leading to a broadened EPR signal, and enabling a complex magnetic state.
With this backdrop of the structure, electronic density-of-states and the magnetic 
structures determined, we now take a look at the magnetism of the three
\SLFCOx\ compounds reflected in experiments.\\

\subsection{\LFCO: Magnetization}
\label{lfco_mag}
\begin{figure}[!t]
	\centering
	\includegraphics[scale=0.55]{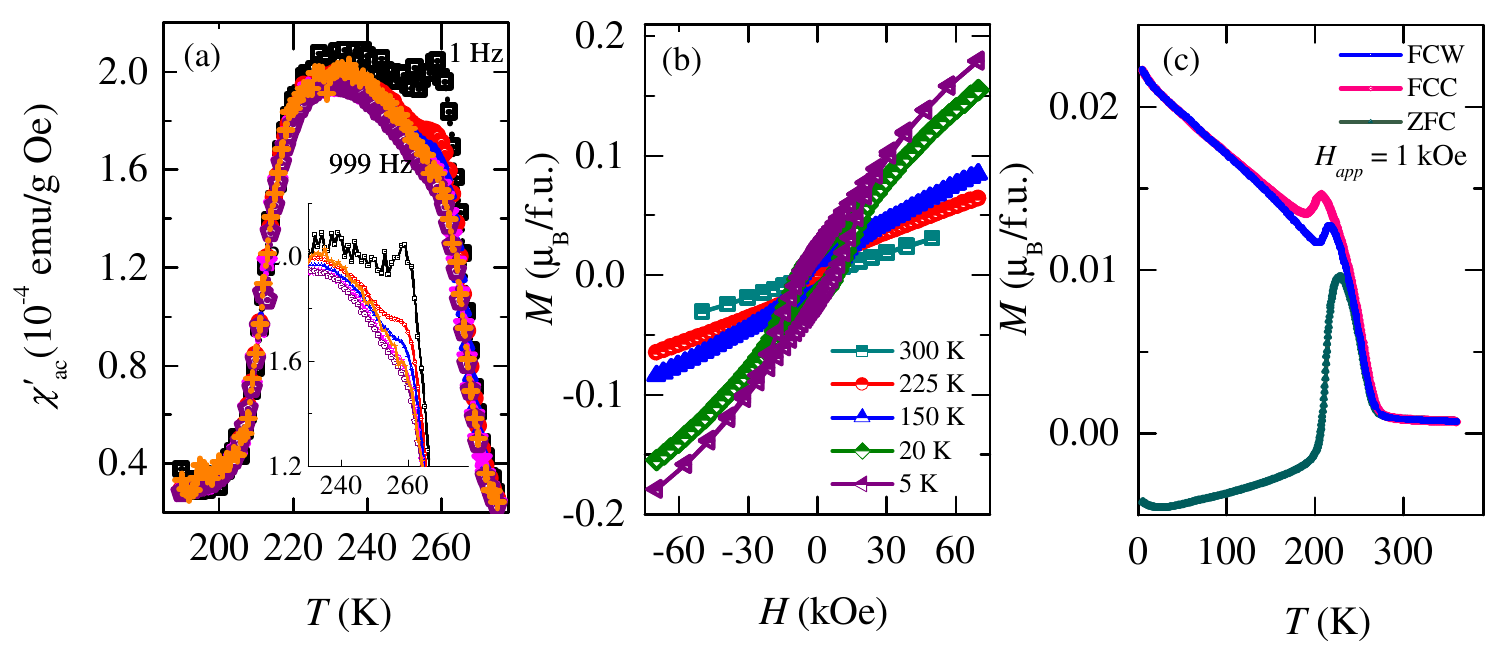}
	\caption{ \label{fig_mag_lfco} (color online) 
		(a) The real part of ac susceptibility, 
		$\chi_\mathrm{ac} (T)$, of \LFCO\ measured 
		at different frequencies in the range, 
		1~Hz $\textendash$ 999~Hz. The phase transitions
		at $T_N$ and $T_S$ are evident. The inset shows 
		a magnified view. (b) The magnetization 
		isotherms at 5~K, 20~K, 150~K, 225~K and 300~K supports
		antiferromagnetism. 
		(c) The dc magnetization FCW and ZFC shows 
		a large bifurcation below $T_S$, where a  
		thermal hysteresis is seen.}
\end{figure}

\indent
Macroscopic magnetization of \LFCO\ measured using ac and dc magnetometry are presented in Fig~\ref{fig_mag_lfco}.
The ac susceptibility, $\chi(T)$, in the frequency range 1~Hz to 999~Hz and temperature 
range 200~K$\textendash$300~K is shown in the panel (a). 
A magnetic phase transition at $T_N \approx$ 270~K is clearly seen in Fig~\ref{fig_mag_lfco} (a). 
A weak frequency dependence of susceptibility is observed at $T_N$. 
In Fig~\ref{fig_mag_lfco} (a), a significant reduction in the magnetization is observed at $T \approx$ 200~K. 
The features in magnetization correlates with the structural phase transition in LFCO from $R\overline{3}c$ to $Pnma$ which is described in detail in the next subsection. 
The isothermal magnetization curves at 5~K, 20~K, 150~K, 225~K and 300~K are 
shown in Fig~\ref{fig_mag_lfco} (b). 
The magnetization isotherms in (b) show hysteresis at low temperatures, especially at 20~K and 5~K. 
However, the maximum magnetic moment attained at 5~K with the application of 7~T is $\mu_\mathrm{max}\approx$ 0.2~$\mu_\mathrm B/$f.u. 
We note that our DFT computations described above are consistent with the macroscopic
magnetization measurements if the magnetic domains are randomly oriented.  
The dc magnetization measurements shown in (c) support the magnetic transition at $T_N$. 
A large irreversibility between the ZFC and FCW curves of magnetization is observed.
Additionally, a strong thermal hysteresis of the FCC and FCW curves is seen around 200~K.
It is revealed later in the next section that it is a structural transition that causes the thermal hysteresis and the large irreversibility.
The magnetic phase transition in the present case occurs close to 300~K while our measurement capability was limited upto 350~K thereby not permitting a
Curie-Weiss analysis in a large paramagnetic range. 

\subsection{\LFCO: Neutron diffraction}
\label{section_neutron}
\begin{figure}[!t]
	\includegraphics[scale=0.45]{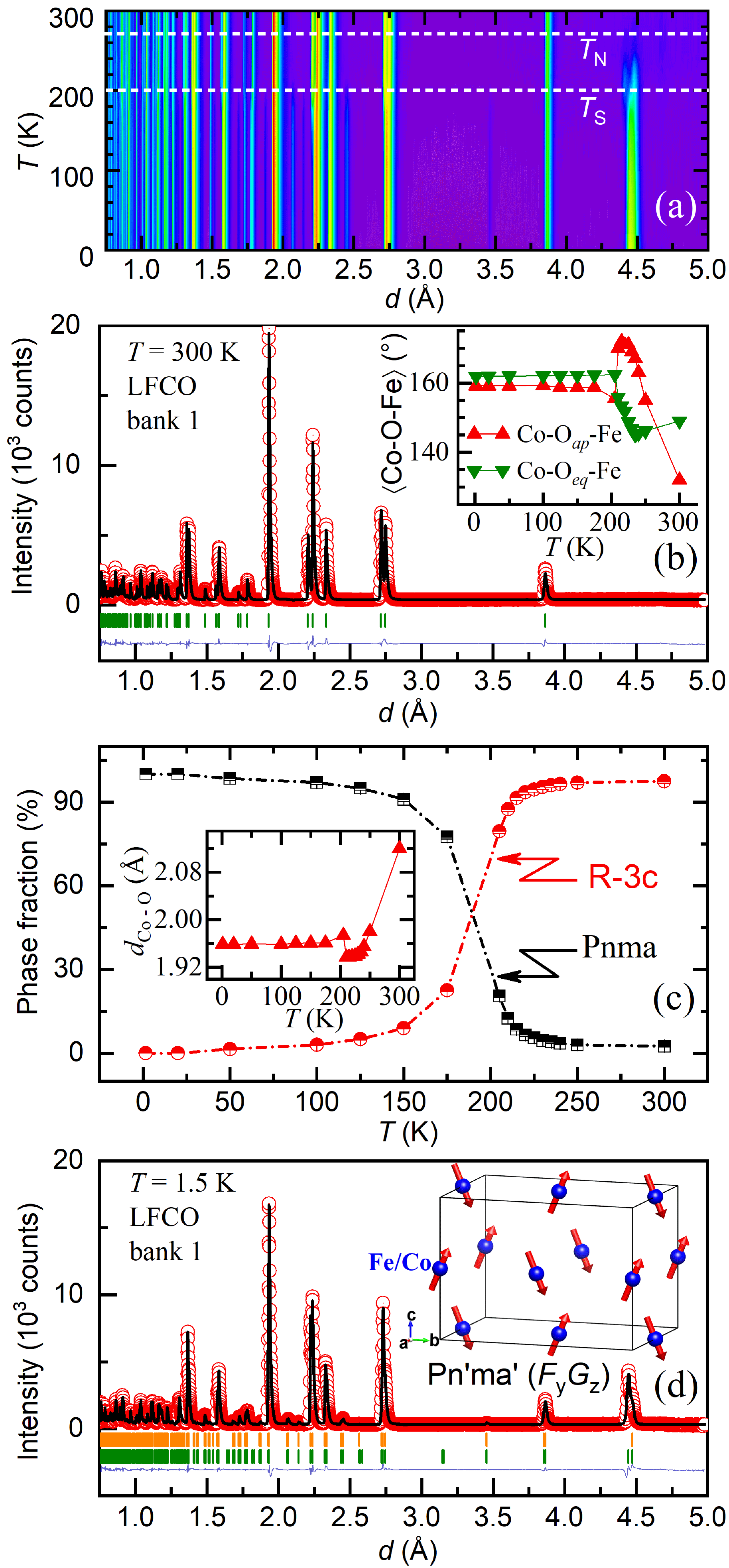}
	\caption{(color online) \label{fig_lfco_npd} (a) A 2D contour of diffracted 
		intensity from \LFCO\ plotted as a function of temperature and $d$-spacing. 
		A magnetic phase transition occurs in LFCO at $T_N \approx$ 270~K 
		and a structural transformation from $R\overline{3}c$ to $Pnma$ 
		at $T_S \approx$ 200~K. (b) The Rietveld refinement of the neutron 
		powder diffraction data at 300~K. The inset shows the temperature 
		variation of bond angles. (c) The percentage distribution of the two 
		structural phases as a function of temperature. The bond distance 
		as a function of temperature is shown in the inset. (d) The Rietveld 
		refinement of the diffraction pattern at 1.5~K where the magnetic 
		structure is faithfully accounted for by the antiferromagnetic 
		$\Gamma_5$ representation (shown in the inset).}
\end{figure}

\indent
In order to understand the magnetic structure of \LFCO\ that would explain the magnetization features observed in Fig~\ref{fig_mag_lfco}, we 
performed neutron diffraction experiments. 
The results are presented in Fig~\ref{fig_lfco_npd}. 
A magnetic anomaly at $T_N \approx$ 270~K can be 
discerned from (a) where the development 
of an additional Bragg peak at $d \approx$ 4.5~{\AA} occurs. 
This feature relates to the (011) and (110) Bragg peaks
	which is indicative of an AFM magnetic structure of the G type.
The Rietveld refinement of the diffraction pattern at 300~K is 
shown in panel (b) where the experimental intensity is plotted in red 
circles and the calculated as black solid line. 
The crystal structure of LFCO at 300~K is refined in the 
rhombohedral space group, $R\overline{3}c$ with the lattice parameters 
$a$ = 5.4935(2)~({\AA}) and $c$ = 13.2343(2)~({\AA}). 
A structural phase transition is observed 
in LFCO at $T_S \approx$ 200~K where the 
crystal structure transforms from $R\overline{3}c$ to orthorhombic $Pnma$. 
Presented in Fig~\ref{fig_lfco_npd} (c) is a plot of the percentage phase 
fraction of the two structural phases as a function of temperature. 
In the intermediate temperature region centered 
around 200~K, mixed structural phases exist. In the inset of 
Fig~\ref{fig_lfco_npd} (b), the bond angle $\langle$Co$\textendash$O$\textendash$Fe$\rangle$ 
and in the inset of (c), the bond distance $d_\mathrm{Fe/Co-O}$ 
in \LFCO\ are shown. Both the bond angles and the bond distances reflect 
strong anomalies around $T_S$ where the structural phase transition occurs. 
The thermal hysteresis in magnetization and the large bifurcation if the 
ZFC/FC curves in LFCO is due to the coexistence of 
mixed $R\overline{3}c$ and $Pnma$ phases over a large temperature range, 
having different magnetization responses to an external magnetic field. \\
\begin{table}[!t]
	\caption{ \label{tab3} The atomic coordinates and lattice 
		parameters of \LFCO\ at 300~K and 1.5~K in $R\overline{3}c$ 
		and $Pnma$ space groups respectively. The structural phase transition 
		to $Pnma$ occurs at $T_S \approx$ 200~K. The lattice parameters 
		at 300~K (for $R\overline{3}c$ ) are $a$ = 5.4935(2)~({\AA}), $c$ = 13.2343(2)~({\AA}) 
		and at 1.5~K (for $Pnma$) are $a$ = 5.4379(5)~({\AA}), $b$ = 7.7055(6)~({\AA}) and 
		$c$ = 5.4886(3)~({\AA}). $W$ stands for Wyckoff position. The goodness-of-fit are
		$\chi^2$ (300~K) = 2.1 and $\chi^2$ (1.5~K) = 2.}
	\setlength{\tabcolsep}{3pt}
	\begin{tabular}{llllllll} \hline\hline
		300~K & $W.$    & $x$ 	        &		$y$		        & 		$z$   \\ \hline
		La & $6a$	& 0	        	&		0       		&		0.25 \\
		Fe & $6b$   & 0		        &		0		        &		0	\\
		Co & $6b$   & 0 		    &		0		        &		0	\\
		O  & $18e$    & 0     		&		0.4461(2)		&		0.25	\\ \hline 
		1.5~K & $W.$     & $x$ 	        &		$y$		        & 		$z$ 	   \\ \hline
		La & $4c$	& 0.0170(9)	   	&		0       		&		0.25	\\
		Fe & $4b$    & 0		        &		0		        &		0.5	\\
		Co & $4b$   & 0 		    &		0		        &		0.5	\\
		O1 & $4c$   & 0.4935(6) 	&		0.25	    	&		0.0631(7)	\\ 
		O2 & $8d$   & 0.2697(6) 	&		0.0385(3)	   	&		0.7304(2)	\\ \hline 
		&   300~K 		& 	1.5~K		\\ \hline
		Co-O$_{ap}$ 	& 	1.9541(12)	&	1.9577(2)		\\
		Co-O$_{eq}$		&	        	&	1.961(4)		\\ 
		Fe-O$_{ap}$ 	& 	1.9541(12)	&	1.9577(2)		\\
		Fe-O$_{eq}$		&	         	&	1.961(4)		\\
		$\langle$ Fe-O$_{ap}$-Co $\rangle$ &   &  159.30(3) \\
		$\langle$ Fe-O$_{eq}$-Co $\rangle$ &   &  160.41(12) \\ \hline\hline
	\end{tabular}
\end{table}

\indent
As the temperature is reduced to 1.5~K, the magnetic Bragg peaks (011) and (110) 
are enhanced in the diffraction pattern, see Fig~\ref{fig_lfco_npd} (d).
This corresponds to the Bragg intensity that develops at $d$ = 4.5~{\AA} at the $T_N$, 
Fig~\ref{fig_lfco_npd} (a). The nuclear structure of LFCO at 1.5~K 
retains the $Pnma$ symmetry. The magnetic structure of LFCO was 
solved after determining the propagation vector through a profile 
fit to the low temperature magnetic peaks ((011) and (110)), thus obtaining 
$\mathrm k$ (0 0 0). The $\mathrm k$-search utility 
within the FullProf Suite was used for this purpose. 
Using this propagation vector, the symmetry-allowed magnetic representations for 
LFCO were determined using SARA$h$ \cite{sarah_wills}.
The crystal structure was assumed to be a pure phase of $Pnma$ in this case and the 
magnetic moments of Fe and Co were assumed to be same since they occupy the 
same crystallographic position within the unit cell. The best description to 
the observed diffraction data was obtained with the $\Gamma_5$ representation
($Pn'ma'$, BNS label 62.448).  A schematic of the 
arrangement of the magnetic moments in the unit cell in $\Gamma_5$ 
representation is shown in the inset of Fig~\ref{fig_lfco_npd} (d), which
shows the $F_yG_z$ AFM structure. During the course of refinement, 
magnetic moment components were allowed to vary along all crystallographic 
directions, however, a negligible value was obtained for the 
$x$-component of the magnetic moment. Absence of a spin re-orientation transition 
at high temperatures was  confirmed and subsequently, the magnetic 
moments were restricted to be in the $y$ and $z$ directions only
in agreement with the DFT calculations. 
After refining the magnetic moments at 4~K, we obtained ordered 
moment of 1.89(7)$\mu_\mathrm{B}/$(Fe,Co) atoms. 
The structural parameters extracted from the
Rietveld refinement of neutron diffraction patterns at 300~K and 1.5~K 
are presented in Table~\ref{tab3}.

\subsection{CHUF magnetization of \SLFCOx}
\label{section_chuf}
\begin{figure}[!t]
	\centering
	\includegraphics[scale=0.34]{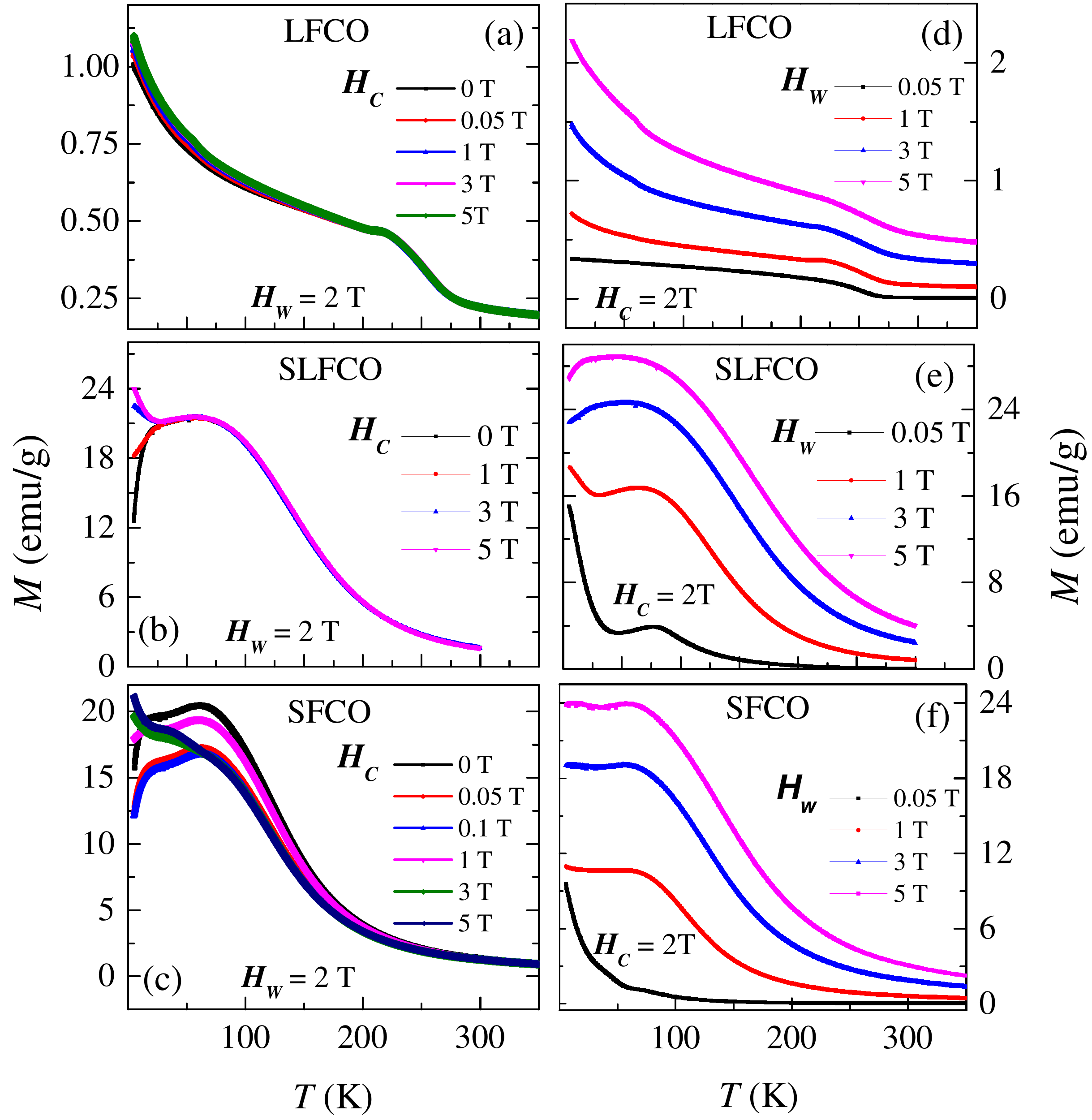}
	\caption{(color online)  \label{fig_chuf} 
		Magnetization curves obtained from performing a CHUF 
		protocol on \LFCO\ (a, d), \SLFCO\ (b, e) and \SFCO\ (c, f). 
		In the panels on the left (a, b, c), a constant 
		warming-field $H_w$ = 2~T was used (CHUF1 protocol) whereas the 
		panels on the right (d, e, f) show curves obtained 
		with a constant cooling-field $H_c$ = 2~T (CHUF2 protocol). 
		Non-equilibrium features are visible in the 
		magnetization of SFCO and SLFCO.}
\end{figure}

From the above sections it is clear that the magnetism of \LFCO\ is different from
the disordered magnetism found in \SFCO\ and \SLFCO \cite{pradheesh,pradheesh2017magneticglass}.
In order to contrast the magnetism in the three compounds,
we performed detailed protocol-based magnetization measurements.
Cooling-and-heating-in-unequal-fields (CHUF) protocol is a 
useful magnetization protocol which can be used to record 
magnetization curves as a function of temperature in order to 
differentiate the non-equilibrium nature of the {\em glass-like} magnetic features
from that of an equilibrium response 
\cite{kushwaha_prb_80_2009low,banerjee_jpcm_21_2009conversion,roy_prb_79_2009contrasting}. 
In what we term as CHUF1 protocol, the sample is cooled across 
the transition temperature in a certain applied magnetic field $H_C$. 
At the lowest temperature, $H_C$ is isothermally changed to 
a different value of measuring field and the magnetization 
is measured while warming the sample. The result of this 
measurement protocol for LFCO, SLFCO and SFCO are 
presented in (a), (b) and (c) respectively in Fig~\ref{fig_chuf}. 
The magnetic field used to measure the magnetization in the warming 
cycle is notated as $H_W$ in the figure. Several values of 
external magnetic fields 0~T, 0.05~T, 0.1~T, 1~T, 3~T and 
5~T were used as $H_C$ to cool the samples (see, (a), (b), (c)). 
In all the three cases, $H_W$ = 2~T was used to measure the magnetization 
while warming. In a second protocol, CHUF2, the cooling field 
$H_C$ was kept constant at 2~T during the time the sample was 
cooled down to low temperature and, subsequently, different fields of 
$H_W$ were used in the warming cycle to measure the magnetization. 
The results of this protocol are presented in (d), (e) and (f) 
of Fig~\ref{fig_chuf} for LFCO, SLFCO and SFCO, respectively. 
In the case of LFCO which orders long-range at high temperature, 
no signature of magnetic relaxation or non-equilibrium dynamics 
is seen in the CHUF1 measurement in (a). 

Note that a $H_C$ upto 5~T 
and a warming field of 2~T does not affect the magnetization features. 
However, apart from a discontinuity in the magnetization 
at $T_N \approx$ 270~K, a second anomaly is discernible at 
$T_S \approx$ 200~K in LFCO, coinciding with 
the structural transformation between $R\overline{3}c$ and 
$Pnma$ phases. The CHUF measurement reveals that LFCO 
behaves in the same way for the $H_C > H_W$ and $H_C < H_W$ 
regimes and hence a magnetic glass-like
state can be ruled out. In the case of CHUF2 protocol, we see 
that the magnetization increases with higher value of measuring 
fields for $H_W$. The anomalies at  270~K 
and 200~K are still present however, with the application of 3~T 
and 5~T for $H_W$, the magnetization at low temperature is enhanced. 
In the case of SLFCO, the CHUF1 protocol shows contrasting effects 
for the two cases, $H_C > H_W$ and $H_C < H_W$, 
as seen in (b). When the cooling field is larger than the measuring 
field, {\em i.e.,} when $H_C > H_W$, the magnetization below the 
anomalous temperature $T_{a1} \approx$ 75~K is significantly increased. 
This observation is consistent with a kinetically arrested ferromagnetic 
state of SLFCO. When SLFCO is warmed up, this glass-like 
arrested ferromagnetic phase reverts to the equilibrium 
antiferromagnetic phase. In the CHUF2 protocol for SLFCO shown 
in the figure panel (e), we can see that the magnetization tends 
to increase in magnitude with higher values of $H_W$. 
Note that there is a drastic difference in the magnetization profile 
below $T_{a1}$ when the CHUF1 and CHUF2 curves of SLFCO are 
compared. From Fig~\ref{fig_chuf} (e) it is clear that when $H_W > H_C$, 
the weak anomaly seen below $T_{a1}$ vanishes and a higher 
magnetization is resulted. The features seen in (b) and (e) 
confirm that SLFCO has a glass-like mixed phase where a large 
volume fraction of the ferromagnetic phase
 devitrifies. In the case of SFCO, the CHUF1 and the CHUF2 data 
 presented in the figures (c) and (f) respectively show signs of 
 magnetic relaxation similar to that of SLFCO albeit weaker in 
 magnitude. The CHUF1 protocol in (c) do indicate that 
the magnetization for $H_C > H_W$ shows an enhanced magnitude 
below the $T_g$.

\subsection{Electron paramagnetic resonance of \SLFCOx}
\label{section_epr}
\begin{figure}[!t]
	\centering
	\includegraphics[scale=0.65]{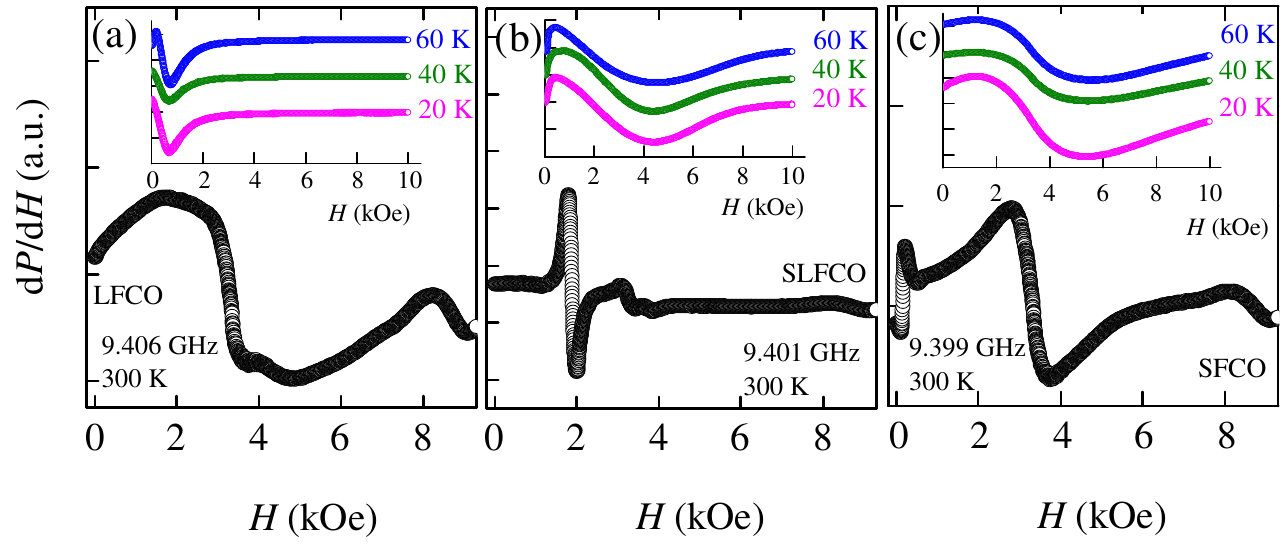}
	\caption{(color online) \label{fig:epr} The electron paramagnetic 
		resonance curves at 300~K for (a) \LFCO, (b) \SLFCO and (c) \SFCO. 
		The inset in each panel show the corresponding curves at low 
		temperatures, 60~K, 40~K and 20~K.
	}
\end{figure}

As another experimental tool to contrast the magnetism
in the three compounds, we use electron paramagnetic resonance (EPR).
The differing features of the magnetic ground states of SFCO, SLFCO and 
LFCO are consistent with the EPR observations presented in Fig~\ref{fig:epr}~(a-c), 
where the EPR signals at 20~K, 40~K, 60~K, and at 300~K are plotted. 
The EPR spectra showed a dramatic dependence on La and Sr 
composition in the present series of compounds, consistent 
with magnetization and neutron diffraction results. 
Figure~\ref{fig:epr} (a) plots the temperature evolution of EPR 
spectrum measured at 20~K, 40~K, 60~K and 300~K for LFCO. 
For LFCO at 300~K, we observe two distinct EPR signals. 
The first signal at $g$ = 2.05(6) (the central field, $H_0$ = 3266~G) 
associated with the peak-to-peak line width ($\Delta H_\mathrm{pp}$) 
of 3266~G, and the second signal appears at $g$ = 0.76(9) ($H_0$ = 8728~G).  
We believe that the former signal is due to the strongly exchange 
coupled Fe$^{3+}$ and Co$^{2+}$ spins, whereas the latter one was 
found to originate from the cavity background, and hence is 
discarded from further discussion. Because of the presence of the two signals, 
a broad Lorentzian curve does not completely account for the EPR 
linshape of LFCO  at 300~K as can be understood from Fig~\ref{fig:epr} (a). 
It can be immediately noticed that as we lower the sample temperature 
from 300~K, a dramatic shift in the EPR signal toward the low field 
region occurs. At 60~K, we detected a complete signal associated 
with $g$ value of 16.07 ($H_0$ = 418~G), characterized by $\Delta H_\mathrm{pp}$ of 552~G. 
These are the benchmark signatures of an ordered antiferromagnetic phase. 

As we started to replace Sr in place of La in LFCO, the 
EPR signal broadens and shifts to the high field region, 
which becomes particularly noticeable at low temperatures 
(b and c panels). Furthermore, the disordered magnetic 
phase increases in abundance upon increasing the Sr content. 
For all the samples, as the temperature increases, the EPR 
signal gets sharper due to motional narrowing effect. Both (b) SLFCO 
and (c) SFCO appear to contain at least two magnetic phases that can give 
rise to spin glass-like behavior, which is consistent with our magnetometry 
results \cite{pradheesh, pradheesh2017magneticglass}. 
The DFT computations also suggested that magnetic multiplets 
are energetically close in these compounds and can coexist at low 
temperatures and also at elevated magnetic fields, thereby 
leading to a broadened EPR signal, enabling a complex magnetic state. 
\begin{table}[!b]
	\caption{\label{tab:epr} The gyromagnetic ratio, $g$, and the 
		linewidth, $\Delta H_\mathrm{pp}$ at different temperatures 
		for \LFCO\ (LFCO) compared with the values for the disordered
		counterparts, \SFCO\ (SFCO) and \SLFCO\ (SLFCO).}
	\setlength{\tabcolsep}{8pt}
	\begin{tabular}{llllll} \hline\hline
		$T$ (K) &    SFCO                 &        LFCO                & SLFCO \\ \hline
		& $g, \Delta H_\mathrm{pp}$ (G) & $g, \Delta H_\mathrm{pp}$ (G)   & $g, \Delta H_\mathrm{pp}$ (G)\\ \hline
		
		60   & 1.90, 4306(2)                & 16.07, 552(1)                     & 2.70, 4110(4)           \\
		
		40   & 2.02, 4053(2)                & 20.41,  --                   & 2.65, 3741(2)           \\  
		
		20   & 2.03, 4111(2)                & 22.02, --                   & 2.75, 3952(2)          \\ \hline\hline
	\end{tabular}
\end{table}

It can be noted that the EPR response of the \SLFCOx\ compounds are 
qualitatively different from that of Sr$_2$FeMoO$_6$ in which strong 
evidence of localized Fe$^{3+}$ cores and itinerant Mo$^{5+}$ electrons 
are found \cite{niebieskikwiat2000high, tovar2002evidence} consistent with
our DFT results that suggest Fe4+. The $g$ 
values and the $H_{pp}$ values estimated from the EPR curves are shown 
in Table~\ref{tab:epr}. We have attempted to fit (not shown) 
the experimental EPR signal to a 
broad Lorentzian function of the form, 
$\frac{dP}{dH}$ = $A \frac{d}{dH}[\Delta H/(4(H − H_0)^2 + \Delta H^2 + \Delta H/(4(H + H_0^2) + \Delta H^2]$, 
where $\Delta H$ is the full-width-at-half-maximum 
which when divided by $\sqrt{3}$ gives the peak-to-peak linewidth 
$\Delta H_{pp}$, and $A$ is proportional to the area under the curve. 
Since the fits were not of high quality due to the presence of 
more than one lineshape terms in the data and also because the 
lineshapes were seen shifted more towards the negative field values
as in the case of LFCO, they are not presented here.

\section{Discussion and conclusions} 
In this section we want to discuss the present results in 
the light of the recent insight we obtained from high-resolution 
inelastic neutron scattering experiments to study the hyperfine 
interactions in SFCO, SLFCO and LFCO 
\cite{chatterji2018hyperfine}.
It was shown that the inelastic signals observed in the two 
structurally and magnetically disordered compounds, 
SFCO and SLFCO were very broad, suggesting a distribution 
of hyperfine fields in these two materials whereas, no inelastic signal 
was observed in the case of LFCO. This suggested no or very 
weak hyperfine field at the Co nucleus due to the Co 
electronic moment. The inelastic spectra of SFCO were 
observed to be significantly narrow which could be attributed to 
a weaker hyperfine local field at the Co nucleus. An assumption 
of heterogeneous local fields at the Co nucleus due to the 
antisite disorder is consistent with SFCO which is a 
spin glass. The results from inelastic studies are in conformity with 
this picture and the model fits to the inelastic spectra suggests a finite 
energy splitting of $\approx$ 1~$\mu$eV (for details of the fits, please 
see Ref [\onlinecite{chatterji2018hyperfine}]).
\\
\indent 
The case of SLFCO appeared interesting as indications of 
electronic spin fluctuations in nano-second time scales were 
observed in the low-$Q$ region, visible in the quasi-elastic channel, 
confirming magnetic short-range order and electronic spin 
freezing below 80~K. 
From the perspective of inelastic neutron scattering, 
the most surprising result was the absence of inelastic 
signal in the ordered state of \LFCO\ down to 1.8~K. 
This implies that the hyperfine field at the Co nucleus for 
this material is extremely weak to measure and that the Co 
moments may not be frozen at very low temperatures.
Thus we surmise that the observed magnetic properties of SFCO, SLFCO
and LFCO are not easily explained solely based on the presence of 
antisite disorder. It is clear that the valence state disorder 
also plays an important role as we observe quasi-elastic scattering 
near the spin freezing temperatures which suggest fluctuations in the 
nanosecond time scale. With the addition of 
Sr in \LFCO,  the spin fluctuations slow down and 
lead to glassy dynamics which is observed through magnetometry.
While it is beyond our computations to address antisite disorder 
directly, our results do suggest that C- and G-structured transition 
metal arrangements are likely to coexist.
\\
\indent
An interesting progression of magnetic ground states is observed in \SLFCOx\ 
as a function of the degree of disorder and with the replacement of La with Sr. 
\LFCO\ has a high temperature magnetic transition 
at $T_N \approx$ 270~K and also a structural phase 
transition at $T_\mathrm{S}\approx$ 200~K where the compound 
transforms from $R\overline{3}c$ to $Pnma$. 
LFCO forms the only magnetically long-range ordered 
member in the series, whereas SFCO and SLFCO are magnetically disordered and form 
respectively, a spin glass and a magnetic glass with a spin freezing 
temperature, $T_\mathrm{g} \approx$ 75~K. 
The structural sensitivity at $\approx$ 200~K in LFCO is reflected in 
the other two compounds SLFCO and SFCO as a weak anomaly in 
the temperature dependence of lattice parameters and the
magnetization. Our neutron diffraction results provide ample 
evidence of magnetic diffuse scattering persisting in SLFCO  upto 300~K.
From the CHUF magnetization protocols, electron paramagnetic 
resonance and neutron diffraction experiments, we mark SLFCO as 
a magnetic glass where nano scale spin fluctuations are evidenced 
through our recent inelastic neutron scattering work. 
Density functional theory calculations performed by adopting the 
crystal structure from the neutron diffraction predicts a AFiM/AFM
ground state which is consistent with the antiferromagnetic 
state arrived at for LFCO through neutron scattering analysis. 
The magnitude of the magnetic moments remained the same as in 
LFCO, however, charge neutrality suggests the presence of mixed 
valence states, in contrast to LFCO and SFCO.
These results align well with the overall picture obtained from our experiments 
for the three compounds from recent inelastic scattering experiments 
where the hyperfine fields of Co was modeled in detail. 
Our present work points toward the importance of competing 
valence state and spin state disorder in realizing different 
magnetic ground states in \SLFCOx\ double perovskites.
Even though our simulation cell (20 atoms) is not large 
enough to address the spin-glass state directly, it provides several insights, 
that distinguish LFCO from SLFCO. Both compositions show strong 
antisite disorder that can support different magnetic signatures. 
While the DFT findings do not provide conclusive evidence for a 
spin-glass state in SFCO and SLFCO, they do suggest that the 
mechanism for spin-glass formation in SLFCO may be facilitated 
by valence state mixing, while in SFCO, it may be attributed to 
coexisting transition metal arrangements and antisite disorder.

\section{Acknowledgements}
HSN acknowledges the UTEP start-up funds and Rising-STARS award
in supporting this work. 
BK would like to acknowledge computing resources
provided through the National Science Foundation (XSEDE)
under grant No. DMR TG-110093.
CMNK acknowledges the financial support by FWF project P27980-N36 
and the European Research Council (ERC Consolidator Grant No 725521).  
SRS, CLS and LMM acknowledge UTEP Start-up funds in supporting this work,  
and National Science Foundation (NSF), USA, with NSF-PREM grant 
DMR-1205302. LMM acknowledges the Wiemer Family for awarding 
Student Endowment for Excellence. The authors thank S. R. J. Hennadige 
for his help in doing EPR measurements.
HGR, PR, KS and VSN acknowledge IITM for funding SVSM.



 \end{document}